\documentclass[conference,letterpaper]{IEEEtran}

\IEEEoverridecommandlockouts

\usepackage{cite}
\usepackage{booktabs}
\usepackage{multirow}
\usepackage{amsmath,amssymb}
\usepackage{mathtools}
\usepackage{amsthm}
\usepackage{algorithm}
\usepackage{algorithmic}
\usepackage{graphicx}
\usepackage{tikz}
\usepackage{pgfplots}
\pgfplotsset{compat=1.18}
\usepgfplotslibrary{groupplots}
\usetikzlibrary{arrows.meta,positioning,shapes.geometric,fit,backgrounds,
                decorations.pathreplacing,calc,matrix,patterns}
\usepackage{xcolor}
\usepackage{subcaption}
\usepackage{tabularx}
\usepackage{makecell}
\usepackage{bm}
\usepackage{url}
\usepackage[T1]{fontenc}
\usepackage{microtype}
\usepackage{enumitem}
\usepackage{balance}
\usepackage{float}
\usepackage{placeins}

\setlist[itemize]{leftmargin=*,topsep=2pt,parsep=1pt,itemsep=2pt}
\setlist[enumerate]{leftmargin=*,topsep=2pt,parsep=1pt,itemsep=2pt}

\definecolor{meshBlue}{RGB}{20,90,160}
\definecolor{meshOrange}{RGB}{215,110,30}
\definecolor{meshGreen}{RGB}{35,140,75}
\definecolor{meshGray}{RGB}{115,115,115}
\definecolor{meshViolet}{RGB}{120,55,160}
\definecolor{meshRed}{RGB}{185,35,35}
\definecolor{lightgray}{RGB}{242,242,242}

\newcommand{\mesh}{\textsc{MeshKV}}
\newcommand{\noc}{\textsc{NoC}}
\newcommand{\kv}{\textsc{KV}}
\newcommand{\takv}{\textsc{TaKV}}
\newcommand{\mare}{\textsc{Mare}}
\newcommand{\pad}{\textsc{Pad}}

\newtheorem{observation}{Observation}

\begin{document}

\title{\mesh: A Network-on-Chip \kv{} Cache Fabric for Scalable Transformer Decoding Accelerators}

\author{%
\IEEEauthorblockN{Dong Liu\textsuperscript{*}}
\IEEEauthorblockA{University of California, Los Angeles\\
Los Angeles, CA, USA\\
\url{pikeliu@ucla.edu}}
\and
\IEEEauthorblockN{Yanxuan Yu\textsuperscript{*}}
\IEEEauthorblockA{Columbia University\\
New York, NY, USA\\
\url{yy3523@columbia.edu}}
}

\maketitle
\renewcommand{\thefootnote}{\fnsymbol{footnote}}%
\footnotetext[1]{Equal contribution.}%
\renewcommand{\thefootnote}{\arabic{footnote}}

\begin{abstract}
Autoregressive transformer decoding is constrained by irregular
key--value~(\kv{}) cache movement on tiled accelerators.
Prior compression and DRAM-placement systems still concentrate traffic on
centralized memory paths that bottleneck long-context serving.
We present \textbf{\mesh}, a \kv{} cache \emph{fabric} that moves blocks
as packetized flows over a lightweight \noc{}.
It co-designs (i)~\textbf{\takv{}} affine striping to spread homes and
cut hotspot load,
(ii)~\textbf{\mare{}} multicast with verified duplicate suppression, and
(iii)~\textbf{\pad{}}, which overlaps prefetch, tile multiply, and streaming
softmax behind credit-aligned FIFOs.
Together they convert bisection back-pressure into useful \kv{} transfer.
On our $8{\times}8$ FPGA implementation with LLaMA-2-7B and Mistral-7B at 8K--32K,
\mesh{} reduces interconnect traffic by up to $\mathbf{58\%}$, improves
\kv{} bandwidth utilization by $\mathbf{2.1\times}$, and delivers up to
$\mathbf{1.9\times}$ multi-stream throughput.
\end{abstract}


\section{Introduction}
\label{sec:intro}

LLM decoding is ``memory-bound,'' but tiled accelerators also stall on
on-chip data movement.
Autoregressive attention scatters \kv{} across a mesh built for dense GEMM,
not all-to-few gather over a growing
prefix~\cite{vaswani2017attention,dao2022flashattention,kwon2023efficient}.

ASICs, FPGAs, and chiplet meshes map compute to a 2-D \textsc{PE} array with
private SRAM.
Prefill amortizes tile hops; decode does not: one query row
($d_h{\approx}128$) meets a \kv{} slab that grows every token.
A centralized design routes these fetches through a shared injection
point, concentrating long-context \kv{} traffic on the same bisection.

\textbf{Our approach.}
We treat the \kv{} cache as a \emph{communication substrate}: each step issues
structured, layer-major reads that admit co-designed placement, routing, and
compute.
\mesh{} is a \noc{}-centric fabric.
Unlike DRAM paging or compression~\cite{kwon2023efficient,sheng2023flexgen},
it optimizes \emph{how} resident SRAM blocks traverse the mesh---multicast from
each home tile, with prefetch overlapping matmul across credit boundaries.

\textbf{Contributions.}
\begin{enumerate}
  \item \textbf{\takv{} (\S\ref{sec:takv}).}
        Affine placement $\pi(\ell,s)$ plus a one-cycle occupancy swap.
  \item \textbf{\mare{} (\S\ref{sec:mare}).}
        Multicast trees, verified duplicate suppression, and a deadlock-free VN0/VN1 split.
  \item \textbf{\pad{} (\S\ref{sec:pad}).}
        Prefetch / multiply / streaming softmax sized to the \noc{} RTT.
  \item \textbf{Evaluation (\S\ref{sec:eval}).}
        Hardware-measured traffic, utilization, throughput, and latency at 8K--32K.
\end{enumerate}

\section{Related Work}
\label{sec:related}

\textbf{KV cache management and serving.}
PagedAttention~\cite{kwon2023efficient}, FlexGen~\cite{sheng2023flexgen},
and TinyServe~\cite{liu2025tinyserve}
virtualize or select \kv{} behind a serving-layer controller; \mesh{} distributes
it \emph{inside} a tiled accelerator with private SRAM.
Quantization and eviction~\cite{liu2024kivi,hooper2024kvquant,zhang2023h2o,li2024snapkv,xiao2023efficient,liu2024llmeasyquant}
shrink the bytes per step; \mesh{} complements these by improving residual
transport.

\textbf{Attention hardware.}
FlashAttention~\cite{dao2022flashattention,dao2023flashattention2} and
FlashDecoding++~\cite{hong2023flashdecoding} tile GPU attention and softmax;
ELSA~\cite{ham2021elsa} and SpAtten~\cite{wang2021spatten} exploit ASIC sparsity.
\mesh{} targets on-chip communication; \pad{} uses online softmax in a
credit-aligned pipeline.

\textbf{Networks-on-chip and FPGA accelerators.}
Classic \noc{}s and the Turn Model~\cite{benini2002networks,dally2004principles,glass1992turn}
supply deadlock theory; Simba~\cite{hu2020simba} uses a 2-D mesh without
\kv{}-class flits.
DFX~\cite{hong2022dfx}, FlightLLM~\cite{zeng2024flightllm}, and
CXL-SpecKV~\cite{liu2026cxl} show HBM/{CXL} \kv{} movement dominates FPGA
decode; \mesh{} replaces point-to-point \kv{} buses with an attention-aware
fabric.

\section{Background and Motivation}
\label{sec:bg}

\subsection{Autoregressive decoding and the bandwidth wall}
\label{sec:bg:decode}

A decoder with $L$ layers, $H$ heads, head dimension $d_h$, and length $T$
computes, for layer $\ell$ and head $h$,
\begin{equation}
  a_{\ell,h} = \mathrm{softmax}\!\left(
      \frac{q_{\ell,h}\,K_{\ell,h,\le t}^{\top}}{\sqrt{d_h}}
    \right),
  \qquad
  o_{\ell,h} = a_{\ell,h}\,V_{\ell,h,\le t},
  \label{eq:attn}
\end{equation}
with $q_{\ell,h} \in \mathbb{R}^{1 \times d_h}$ and cached slabs
$K_{\ell,h,\le t}, V_{\ell,h,\le t} \in \mathbb{R}^{t \times d_h}$.
The cache occupies
\begin{equation}
  \Gamma(T) = 2 L H d_h T \cdot \mathrm{bpe} \;\text{ bytes},
  \label{eq:kvcache}
\end{equation}
with $\mathrm{bpe}{=}2$ for FP16: ${\approx}$\,16\,GiB for LLaMA-2-7B
($L{=}32$, $H{=}32$, $d_h{=}128$) at $T{=}32\,768$, and every decode step
\emph{reads all of it}.
FlashAttention~\cite{dao2022flashattention,dao2023flashattention2} reduces the
constant for batched prefill via tiled SRAM reuse, but $B{=}1$ decode exposes
every byte with no reuse.

\subsection{Tiled accelerators and the spoke--hub pathology}
\label{sec:bg:hub}

\begin{figure}[t]
\centering
\begin{tikzpicture}
\begin{axis}[
  width=0.44\textwidth,
  height=4.4cm,
  ybar stacked,
  bar width=14pt,
  ymin=0, ymax=100,
  ylabel={Link utilization (\%)},
  xtick={1,2,3},
  xticklabels={\textsc{Cent}, \textsc{Shared}, \mesh},
  enlarge x limits=0.35,
  legend style={at={(0.5,1.02)},anchor=south,legend columns=3,font=\scriptsize},
  grid=major, grid style={dashed},
  every axis plot/.append style={draw=none},
]
\addplot[fill=meshBlue!70] coordinates {(1,19) (2,29) (3,61)};
\addplot[fill=meshOrange!80] coordinates {(1,41) (2,30) (3,11)};
\addplot[fill=meshGray!60] coordinates {(1,40) (2,41) (3,28)};
\legend{Useful \kv{} transfer, Back-pressure stall, Other}
\end{axis}
\end{tikzpicture}
\caption{Breakdown of link-cycle utilization at the mesh bisection
  ($T{=}32\mathrm{K}$, LLaMA-2-7B, $B{=}1$).
  \textsc{Cent} expends $41\%$ of bandwidth on back-pressure;
  \mesh{} converts that capacity into useful \kv{} transfer.}
\label{fig:motiv-traffic}
\end{figure}
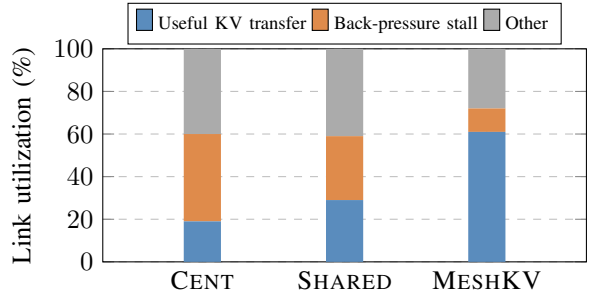

Tile-based accelerators (e.g., Simba~\cite{hu2020simba}, custom chiplets)
map heads and layers across a 2-D \textsc{PE} mesh.
Prefill carries dense GEMM traffic; decode issues width-$d_h$ gathers from
wherever a \kv{} block lives to wherever the compute tile sits.
When \kv{} sits at one logical memory controller, all such reads converge at
the same injection point.
The mesh degenerates to a spoke--hub: $H$ requesters fan in to one hub,
consuming 40--60\% of link cycles on back-pressure
(Figure~\ref{fig:motiv-traffic}).

\subsection{Observed locality in \kv{} accesses}
\label{sec:bg:locality}

An instrumented FP16 decoder on LLaMA-2-7B~\cite{touvron2023llama2} at
$T \in \{8192, 32768\}$ shows two regularities that motivate \mesh{}:

\begin{observation}[Layer-major temporal locality]
\label{obs:layermajor}
Within a decode step, layers are visited in order $\ell = 0, \ldots, L{-}1$;
a \kv{} line is reused only at step $t{+}1$, so the prefetch window is
exactly $L$ layer traversals.
\end{observation}

\begin{observation}[Head-group spatial locality]
\label{obs:headgroup}
Heads that share a prefix span request identical segments
(Mistral-7B GQA uses $G{=}4$; LLaMA-2-7B is multi-head).
At $T{=}32\mathrm{K}$, $78\%$ of segment requests arrive within 12 cycles at
the same router, so multicast coalescing does not require wide merge buffers.
\end{observation}

These observations underlie \takv{} placement (\S\ref{sec:takv}) and
\mare{} coalescing (\S\ref{sec:mare}): \textsc{KV\_DATA} fans out, while
\textsc{PART} converges.

%

\section{System Architecture Overview}
\label{sec:overview}

\begin{figure}[t]
\centering
\begin{tikzpicture}[
  tile/.style={draw=meshBlue!40, rounded corners=3pt, fill=white,
               minimum width=1.72cm, minimum height=1.72cm, line width=0.80pt},
  macarr/.style={draw=meshBlue!70, fill=meshBlue!15, rounded corners=1.5pt,
                 minimum width=0.58cm, minimum height=0.38cm,
                 font=\fontsize{3.8}{4.5}\selectfont, align=center, inner sep=0.8pt},
  kvsram/.style={draw=meshGreen!70, fill=meshGreen!15, rounded corners=1.5pt,
                 minimum width=0.58cm, minimum height=0.55cm,
                 font=\fontsize{3.8}{4.5}\selectfont, align=center, inner sep=0.8pt},
  scrpad/.style={draw=meshViolet!70, fill=meshViolet!13, rounded corners=1.5pt,
                 minimum width=0.58cm, minimum height=0.24cm,
                 font=\fontsize{3.2}{3.8}\selectfont, align=center, inner sep=0.5pt},
  rtr/.style={draw=meshOrange!78, fill=meshOrange!22, rounded corners=2pt,
              minimum width=0.36cm, minimum height=0.36cm,
              font=\fontsize{3.8}{4.5}\selectfont, align=center, inner sep=0.4pt,
              line width=0.75pt},
  lnk/.style={meshGray!50, line width=0.62pt},
  lnk_hot/.style={meshOrange!75, line width=1.55pt},
  brc/.style={meshOrange!95, line width=1.65pt, -{Stealth[length=3.2pt,width=2.3pt]}},
  ucast/.style={meshBlue!75, line width=1.0pt, dashed, -{Stealth[length=2.6pt,width=1.8pt]}},
  pref/.style={meshRed!55, line width=0.7pt, dotted, -{Stealth[length=2pt,width=1.4pt]}},
  dcbar/.style={draw=meshViolet!55, fill=meshViolet!10, rounded corners=1.5pt,
                minimum width=1.62cm, minimum height=0.19cm,
                font=\fontsize{3.5}{4}\selectfont, align=center, inner sep=0.6pt},
  annot/.style={draw=meshGray!42, fill=white, rounded corners=1.5pt,
                font=\fontsize{3.8}{4.6}\selectfont, inner sep=1.6pt, align=left},
]
\def\sp{2.10}

\begin{scope}[on background layer]
  \fill[meshBlue!6, rounded corners=4pt]
    ({-0.64},{-0.60}) rectangle ({1.5*\sp-0.05},{3*\sp+0.60});
  \fill[meshGreen!6, rounded corners=4pt]
    ({1.5*\sp+0.05},{-0.60}) rectangle ({3*\sp+0.74},{3*\sp+0.60});
  \fill[meshOrange!13, rounded corners=3pt]
    ({1*\sp-0.87},{1*\sp-0.87}) rectangle ({1*\sp+0.87},{1*\sp+0.87});
\end{scope}

\foreach \col in {0,1,2,3}{
  \foreach \row in {0,1,2,3}{
    \node[tile] (T\col\row) at ({\col*\sp},{\row*\sp}) {};
    \node[macarr] at ([xshift=-0.33cm,yshift=+0.34cm]T\col\row) {MAC\\arr};
    \node[kvsram] at ([xshift=+0.36cm,yshift=+0.22cm]T\col\row) {KV\\SRAM};
    \node[scrpad] at ([xshift=-0.33cm,yshift=-0.37cm]T\col\row) {Scrpad};
    \node[rtr] (R\col\row) at ([xshift=+0.36cm,yshift=-0.36cm]T\col\row) {\textbf{R}};
    \draw[meshGray!22, line width=0.38pt]
      ([xshift=+0.01cm,yshift=-0.79cm]T\col\row)
      -- ([xshift=+0.01cm,yshift=+0.79cm]T\col\row);
    \draw[meshGreen!40, line width=0.33pt]
      ([xshift=+0.08cm,yshift=+0.04cm]T\col\row)
      -- ([xshift=+0.63cm,yshift=+0.04cm]T\col\row);
    \draw[meshGreen!40, line width=0.33pt]
      ([xshift=+0.08cm,yshift=+0.22cm]T\col\row)
      -- ([xshift=+0.63cm,yshift=+0.22cm]T\col\row);
    \node[font=\fontsize{3.0}{3.5}\selectfont, meshGray!48,
          anchor=north west, inner sep=0pt]
      at ([xshift=-0.83cm,yshift=+0.83cm]T\col\row) {(\col,\row)};
  }
}

\draw[meshOrange!68, line width=1.35pt, rounded corners=3pt]
  ({1*\sp-0.87},{1*\sp-0.87}) rectangle ({1*\sp+0.87},{1*\sp+0.87});

\foreach \col in {0,1,2}{
  \foreach \row in {0,1,2,3}{
    \pgfmathtruncatemacro{\nc}{\col+1}
    \draw[lnk] (R\col\row) -- (R\nc\row);
  }
}
\foreach \col in {0,1,2,3}{
  \foreach \row in {0,1,2}{
    \pgfmathtruncatemacro{\nr}{\row+1}
    \draw[lnk] (R\col\row) -- (R\col\nr);
  }
}
\draw[lnk_hot] (R11) -- (R21);
\draw[lnk_hot] (R21) -- (R31);
\draw[lnk_hot] (R21) -- (R22);

\draw[meshOrange!88, line width=1.75pt]
  ({1.5*\sp},{-0.60}) -- ({1.5*\sp},{3*\sp+0.60})
  node[above, font=\scriptsize\bfseries, meshOrange!95] {\textbf{bisection}};

\foreach \col in {0,1,2,3}{
  \node[dcbar] (DC\col) at ({\col*\sp},{3*\sp+1.10}) {DC$_{\col}$};
  \draw[meshViolet!40, line width=0.50pt, -{Stealth[length=2pt,width=1.4pt]}]
    (DC\col.south) -- (T\col3.north);
}

\node[font=\scriptsize\bfseries, meshBlue!62]  at ({0.5*\sp},{3*\sp+1.42}) {Q2};
\node[font=\scriptsize\bfseries, meshBlue!62]  at ({0.5*\sp},{-0.46})       {Q0};
\node[font=\scriptsize\bfseries, meshGreen!62] at ({2.5*\sp},{3*\sp+1.42}) {Q3};
\node[font=\scriptsize\bfseries, meshGreen!62] at ({2.5*\sp},{-0.46})       {Q1};

\draw[brc] (R11) -- (R21)
  node[midway, above=2pt, font=\fontsize{4.5}{5.2}\selectfont, meshOrange]
  {\textsc{kv\_data}};
\draw[brc] (R21) -- (R31);
\draw[brc] (R21) -- (R22);

\node[draw=meshOrange!55, fill=meshOrange!10, rounded corners=1.2pt,
      font=\fontsize{3.2}{3.8}\selectfont, inner sep=1pt, align=center,
      anchor=south west] at ([xshift=0.04cm,yshift=0.18cm]R21)
  {mcast\\branch};

\draw[ucast] (R21) to[bend right=30]
  node[midway, right=2pt, font=\fontsize{4.5}{5.2}\selectfont, meshBlue!82]
  {\textsc{part}} (R20);

\node[draw=meshOrange!58, fill=meshOrange!12, rounded corners=1.5pt,
      font=\fontsize{3.5}{4.2}\selectfont, inner sep=1.2pt, align=center,
      anchor=north] at ({1*\sp},{1*\sp-0.92})
  {\textsf{home}: $\pi(\ell,s){=}(1,1)$};

\node[draw=meshGreen!62, fill=meshGreen!14, rounded corners=1pt,
      font=\fontsize{3.0}{3.6}\selectfont, inner sep=0.9pt, align=center,
      anchor=south east] at ([xshift=-0.02cm,yshift=+0.17cm]R21)
  {BF};

\draw[{Stealth[length=2pt,width=1.4pt]}-{Stealth[length=2pt,width=1.4pt]},
      meshRed!62, line width=0.85pt]
  ({1.5*\sp-0.12},{0.45*\sp}) -- ({1.5*\sp+0.12},{0.45*\sp});
\node[font=\fontsize{3.2}{3.8}\selectfont, meshRed!68, anchor=north]
  at ({1.5*\sp},{0.45*\sp-0.04}) {\textsc{Pf} credit};

\node[draw=meshRed!48, fill=meshRed!8, rounded corners=2pt,
      font=\fontsize{4.5}{5.5}\selectfont, align=center, inner sep=2pt]
  (hbmlbl) at ({1.5*\sp},{-1.32})
  {HBM2 · 8\,GB · 460\,GB/s};
\foreach \col in {0,1,2,3}{
  \pgfmathsetmacro{\xoff}{(\col-1.5)*0.6*\sp}
  \draw[pref] (T\col0.south)
    to[out=270,in=90] ([xshift=\xoff cm]hbmlbl.north);
}
\node[font=\fontsize{3.5}{4.2}\selectfont, meshRed!60, anchor=north]
  at ({1.5*\sp},{-1.60}) {tile-local prefetch descriptors};

\node[font=\fontsize{3.8}{4.6}\selectfont, meshGreen!72, align=left,
      anchor=north west] at ({3*\sp+0.78},{3*\sp+0.55})
  {\textit{memory}\\[-1pt]\textit{plane}\\[-1pt](\takv)};
\node[font=\fontsize{3.8}{4.6}\selectfont, meshOrange!78, align=left,
      anchor=west] at ({3*\sp+0.78},{1.5*\sp})
  {\textit{router}\\[-1pt]\textit{plane}\\[-1pt](\mare)};
\node[font=\fontsize{3.8}{4.6}\selectfont, meshBlue!72, align=left,
      anchor=south west] at ({3*\sp+0.78},{-0.55})
  {\textit{compute}\\[-1pt]\textit{plane}\\[-1pt](\pad)};

\node[annot, anchor=south west] at ({-0.62},{-1.60}) {%
  {\color{meshOrange!90}\rule[1pt]{7pt}{1.8pt}}\;VN1: \textsc{kv\_data} multicast\\[1.5pt]
  {\color{meshBlue!75}\rule[1pt]{3pt}{1pt}}\,{\color{meshBlue!75}\rule[1pt]{3pt}{1pt}}\;VN0: \textsc{part} unicast\\[1.5pt]
  {\color{meshRed!60}\dotfill\rule[1pt]{0pt}{0pt}}\;HBM prefetch};

\node[draw=meshGray!38, fill=white, rounded corners=2pt,
      inner sep=2.5pt, font=\fontsize{5}{6}\selectfont,
      anchor=north east] at ({3*\sp+0.72},{-0.64}) {%
  \textcolor{meshBlue!80}{\rule[1pt]{5pt}{5pt}}\;MAC\enspace
  \textcolor{meshGreen!80}{\rule[1pt]{5pt}{5pt}}\;KV SRAM\enspace
  \textcolor{meshViolet!80}{\rule[1pt]{5pt}{5pt}}\;Scrpad\enspace
  \textcolor{meshOrange!80}{\rule[1pt]{5pt}{5pt}}\;Router};

\end{tikzpicture}
\caption{$4{\times}4$ \mesh{} tile array (scalable to $8{\times}8$).
  Home tile $(1,1)$ roots a \mare{} multicast for \textsc{KV\_DATA}; VN0
  carries \textsc{PART} reduction; \pad{} prefetches from HBM.}
\label{fig:arch-overview}
\end{figure}
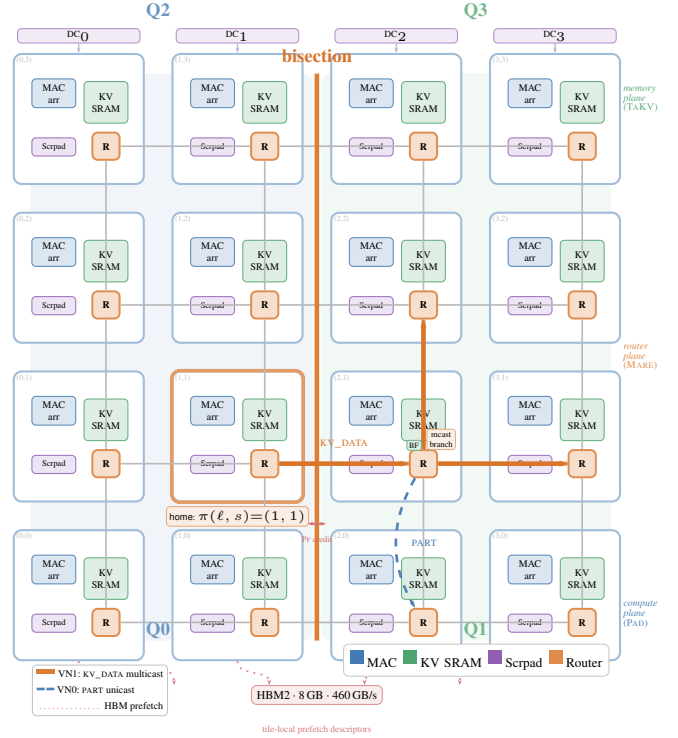

Figure~\ref{fig:arch-overview} deploys \mesh{} on a $4{\times}4$ mesh of
tiles $(x,y)\in\{0,\ldots,W{-}1\}{\times}\{0,\ldots,H{-}1\}$
(scalable to $8{\times}8$).
Each tile has a MAC array, four-bank \kv{} SRAM, a partial-sum scratchpad, and a
two-VN router.
A per-column controller broadcasts descriptors $(\ell,h)$; tiles resolve
$\pi(\ell,s)$ and inject \textsc{KV\_FETCH} on VN1.
The controller is off the critical path (4-cycle combinational \takv{}).
Four-bank \kv{} SRAM lets a tile stream the next window while MACs consume
the current flits.
A decode step broadcasts descriptors, resolves $\pi(\ell,s)$, injects
\textsc{KV\_FETCH} on VN1, receives \textsc{KV\_DATA} at the compute tile,
and returns \textsc{PART} on VN0 to a column reduce root.

Three planes: \takv{} places $\mathcal{B}(\ell,s)$ at load time (with lazy
swaps); \mare{} coalesces, multicasts, and arbitrates VCs; \pad{} runs
\textsc{Pf}/\textsc{Tm}/\textsc{Rd} and credits the router.
Multicast trees are therefore rooted at the home tile $\pi(\ell,s)$, not at a
shared memory port.

\section{Topology-Aware \kv{} Striping (\takv{})}
\label{sec:takv}

Macro-block $\mathcal{B}(\ell,s)$ covers $P$ consecutive tokens of layer $\ell$
across all heads and both $K$ and $V$:
\begin{equation}
  |\mathcal{B}(\ell,s)| = 2 H d_h P \cdot \mathrm{bpe}
  \quad\text{bytes},\qquad
  s = \lfloor j / P \rfloor.
  \label{eq:block}
\end{equation}
With $P{=}64$, $H{=}32$, $d_h{=}128$, $\mathrm{bpe}{=}2$, $|\mathcal{B}|=1\,\mathrm{MB}$
is the placement and multicast unit.

On a $W{\times}H$ mesh, \takv{} maps $\mathcal{B}(\ell,s)$ to
\begin{equation}
  \pi(\ell,s) = \bigl(\ell \bmod W,\; (a\cdot\ell + s)\bmod H\bigr),
  \label{eq:pi}
\end{equation}
with $a$ odd and $\gcd(a,H)=1$ (e.g.\ $a = W{+}1$).
Algorithm~\ref{alg:takv} is four combinational instructions.

\begin{algorithm}[t]
\caption{\takv{} static placement.}
\label{alg:takv}
\begin{algorithmic}[1]
\REQUIRE Layer $\ell$, segment $s$, mesh dimensions $W, H$, constant $a$ (odd, coprime to $H$)
\ENSURE Home-tile coordinates $(x, y)$
\STATE $x \;\gets\; \ell \bmod W$
\STATE $y \;\gets\; (a \cdot \ell + s) \bmod H$
\STATE \textbf{return} $(x, y)$
\end{algorithmic}
\end{algorithm}

Affine $\pi(\ell,s)$ does not shorten expected distance to a
\emph{fixed} home: on a symmetric mesh a uniform requester has the same mean
hops to any single destination.
The gain is spreading homes so injection and bisection load
$\max_e \lambda_e$ drop, converting the back-pressure stall in
Figure~\ref{fig:motiv-traffic} into useful \kv{} transfer.
The coprime stride $a{=}W{+}1$ permutes the $y$ coordinate as $\ell$
increments, so consecutive layers occupy different columns and consecutive
segments of one layer walk distinct rows.
Injection therefore spreads even though mean hops to any single home do not
shrink.
Occupancy swaps further flatten $\sum_e \bar{q}_e^2$.

A $W{\times}H$ mesh assigns at most
$N_{\max} = \bigl\lceil L\lceil T/P\rceil/(WH)\bigr\rceil$ blocks per tile
($256$\,MB of logical \kv{} at $L{=}32$, $T{=}32768$, $P{=}64$, $W{=}H{=}8$).
U280 provides HBM2 ($8$\,GB, $460$\,GB/s) and DDR4 ($32$\,GB); the $16$\,GiB
slab does not fit in HBM alone.
Table~\ref{tab:fpga}'s $512$ URAMs give eight URAMs ($288$\,KB) per tile---a
four-bank streaming window, not a $1$\,MB block---so homes promote HBM/DDR
data into URAM under tile-local prefetch descriptors.

Every $K{=}128$ steps (one SRAM eviction epoch), routers sample occupancy
$\bar{q}_e$ from existing credit counters and swap two homes iff the swap
strictly reduces $\sum_e \bar{q}_e^2$ and preserves per-tile capacity
$N_{\max}$
(4 comparisons, 2 register writes, one control cycle).

\section{Multicast Routing Engine (\mare{})}
\label{sec:mare}

\mesh{} classifies on-chip traffic into three flit types
(Table~\ref{tab:flit}): \textsc{KV\_FETCH} headers
$(\mathrm{seg},\ell,\mathrm{mask},\mathrm{coord})$ and \textsc{KV\_DATA}
bodies (512-bit payload, dedup tag) on VN1 decouple fetch coalescence from
data fan-out; \textsc{PART} partial logits use VN0.
At the home tile $\pi(\ell,s)$, requests arriving in window
$W_{\mathrm{coal}}{=}12$ cycles merge head masks
\begin{equation}
  M(\ell,s)=\bigvee_{r\in W_{\mathrm{coal}}}\mathrm{mask}_r,
  \label{eq:mask}
\end{equation}
then an XYM spanning tree over destinations $\mathcal{D}(M)$ multicasts each
payload once per branch ($\mathcal{O}(H\log H)$ build, $H{=}32$, $+4$ control
cycles before \textsc{KV\_DATA} departs).
Replication occurs only at tree branches, so a payload crosses any link at
most once; the exact-tag plus bloom path is a safety net for overlapping
trees and late arrivals.
Each router keeps a 16-entry exact tag table of in-flight $(\mathrm{seg},\ell)$
and a $256$-bit bloom as a prefilter over $W_b{=}96$ cycles.
A duplicate is suppressed only on an exact-table hit; a miss or a full table
forwards, so bloom false positives never drop a first-arrival payload
($32$\,B bloom plus 16 tags).

\begin{table}[t]
\centering
\caption{Flit classes in \mesh{}.}
\label{tab:flit}
\small
\begin{tabularx}{\columnwidth}{@{}l c c X@{}}
\toprule
\textbf{Class} & \textbf{Dir.} & \textbf{VC} & \textbf{Contents} \\
\midrule
\textsc{KV\_FETCH} & requester$\to$home  & VN1 &
  seg id, layer, head mask (32 bits), requester coord \\
\textsc{KV\_DATA}  & home$\to$requester  & VN1 &
  512-bit \kv{} payload, in-flight segment tag \\
\textsc{PART}      & tile$\to$reduce root & VN0 &
  512-bit partial logit accumulator, softmax running stats \\
\bottomrule
\end{tabularx}
\end{table}

\subsection{Deadlock avoidance}
\label{sec:mare:deadlock}

\mesh{} splits traffic across two virtual networks: VN0 carries unicast
\textsc{PART} reductions on dimension-ordered \textsc{XY} routing, while VN1
carries \textsc{KV\_FETCH} and \textsc{KV\_DATA} multicast on XYM-restricted
turns~\cite{glass1992turn}.
No flit holds a VN0 buffer while waiting for VN1 resources or vice versa, so
inter-network cycles are structurally eliminated; restricted-turn routing
removes intra-network cycles within each VN.

\section{Pipelined Attention Dataflow (\pad{})}
\label{sec:pad}

\begin{figure}[t]
\centering
\begin{tikzpicture}[
  stg/.style={draw, rounded corners=3pt,
              minimum width=3.90cm, minimum height=0.88cm,
              align=center, font=\scriptsize, inner sep=3pt, line width=0.82pt},
  fifo/.style={draw=meshGray!55, fill=meshGray!9, rounded corners=2pt,
               minimum width=1.55cm, minimum height=0.37cm,
               font=\fontsize{4.2}{5}\selectfont, align=center, inner sep=1.5pt,
               line width=0.65pt},
  darr/.style={-{Stealth[length=2.8pt,width=2pt]}, line width=0.88pt, meshGray!62},
  narr/.style={-{Stealth[length=2.6pt,width=1.9pt]}, line width=0.98pt},
  tcell/.style={draw, minimum width=0.90cm, minimum height=0.34cm,
                align=center, font=\fontsize{3.5}{4.2}\selectfont, inner sep=0.8pt},
]

\node[font=\fontsize{3.8}{4.6}\selectfont, meshBlue!72, anchor=east]
  at (-2.12, 1.76) {\textsc{Pf}:};
\node[font=\fontsize{3.8}{4.6}\selectfont, meshOrange!78, anchor=east]
  at (-2.12, 1.36) {\textsc{Tm}:};
\node[font=\fontsize{3.8}{4.6}\selectfont, meshGreen!72, anchor=east]
  at (-2.12, 0.96) {\textsc{Rd}:};

\foreach \i/\lbl in {0/$s$, 1/$s{+}1$, 2/$s{+}2$, 3/$s{+}3$}{
  \node[tcell, fill=meshBlue!22, draw=meshBlue!52]
    at ({-1.65 + \i*0.92}, 1.76) {\lbl};
}
\foreach \i/\lbl in {0/$s{-}1$, 1/$s$, 2/$s{+}1$, 3/$s{+}2$}{
  \node[tcell, fill=meshOrange!22, draw=meshOrange!52]
    at ({-1.65 + \i*0.92}, 1.36) {\lbl};
}
\foreach \i/\lbl in {0/$s{-}2$, 1/$s{-}1$, 2/$s$, 3/$s{+}1$}{
  \node[tcell, fill=meshGreen!18, draw=meshGreen!48]
    at ({-1.65 + \i*0.92}, 0.96) {\lbl};
}
\draw[-{Stealth[length=2.4pt,width=1.7pt]}, meshGray!50, line width=0.65pt]
  (-2.10, 0.73) -- (1.58, 0.73)
  node[right, font=\fontsize{3.5}{4.2}\selectfont, meshGray!62] {time $\to$};

\node[stg, fill=meshBlue!14, draw=meshBlue!62] (pf) at (0, 0) {%
  \textbf{Stage~0:}\;\textsc{Pf}\quad\textit{prefetch}\\[-0.5pt]
  issue \textsc{kv\_fetch}$(s{+}1)$\;$\cdot$\;credit counter};

\draw[darr] (pf.south) -- (0,-0.62);
\node[fifo] (f01) at (0,-0.81) {async FIFO\;$F{=}48$\,flits};
\draw[darr] (f01.south) -- (0,-1.18);

\node[stg, fill=meshOrange!14, draw=meshOrange!62] (tm) at (0,-1.62) {%
  \textbf{Stage~1:}\;\textsc{Tm}\quad\textit{tile multiply}\\[-0.5pt]
  $q_{\ell,h}\cdot K_s$;\;logit accumulate in scratchpad};

\draw[darr] (tm.south) -- (0,-2.43);
\node[fifo] (f12) at (0,-2.62) {async FIFO\;$F{=}48$\,flits};
\draw[darr] (f12.south) -- (0,-2.99);

\node[stg, fill=meshGreen!13, draw=meshGreen!58] (rd) at (0,-3.24) {%
  \textbf{Stage~2:}\;\textsc{Rd}\quad\textit{stream reduce}\\[-0.5pt]
  online softmax:\;update $(m_s,\ell_s,\mathbf{o}_s)$};

\draw[darr] (rd.south) -- ++(0,-0.30)
  node[below, font=\fontsize{4.5}{5.5}\selectfont, meshGreen!72, align=center]
  {$\mathbf{o}_{\mathrm{layer}}$ (FP16)};

\draw[narr, meshOrange!82]
  (pf.east) -- ++(0.88, 0)
  node[right, font=\fontsize{3.8}{4.6}\selectfont, meshOrange!90, align=left]
  {\textsc{kv\_fetch}\\[-1pt]{\fontsize{3.2}{3.8}\selectfont\color{meshGray!65}VN1 $\to$ home tile}};

\draw[narr, meshBlue!72]
  ($(f01.east)+(0.88,0)$) -- (f01.east)
  node[pos=0.0, right, font=\fontsize{3.8}{4.6}\selectfont, meshBlue!82, align=left]
  {\textsc{kv\_data}\\[-1pt]{\fontsize{3.2}{3.8}\selectfont\color{meshGray!65}from home tile}};

\draw[{Stealth[length=2pt,width=1.4pt]}-{Stealth[length=2pt,width=1.4pt]},
      meshRed!55, line width=0.82pt]
  (3.06, 0) -- (3.06,-0.81);
\node[font=\fontsize{3.5}{4.2}\selectfont, meshRed!62, anchor=west, align=left]
  at (3.14,-0.40) {RTT\\$2L_{\max}{+}t_{\mathrm{ack}}$\\${\approx}22$\,cy};

\node[draw=meshGray!38, dashed, rounded corners=4pt, line width=0.72pt,
      fit=(pf)(f01)(tm)(f12)(rd), inner sep=5.5pt] (onelayer) {};
\node[font=\fontsize{4.5}{5.5}\selectfont, meshGray!60, anchor=north east]
  at (onelayer.south east) {$\leftarrow$\,one layer};

\draw[decorate, decoration={brace, amplitude=3.5pt, mirror},
      meshBlue!60, line width=0.78pt]
  (-2.22, 0.46) -- (-2.22,-3.68)
  node[midway, left=4pt, font=\fontsize{4.0}{5.0}\selectfont, align=right, meshBlue!68]
  {overlap\\across\\segments};

\end{tikzpicture}
\caption{\pad{} three-stage pipeline (\textsc{Pf}/\textsc{Tm}/\textsc{Rd}).
  Top: segment overlap conceals \noc{} RTT (${\approx}22$ cycles).
  Bottom: \kv{} prefetch, tile multiply, and online softmax.}
\label{fig:pad-pipeline}
\end{figure}
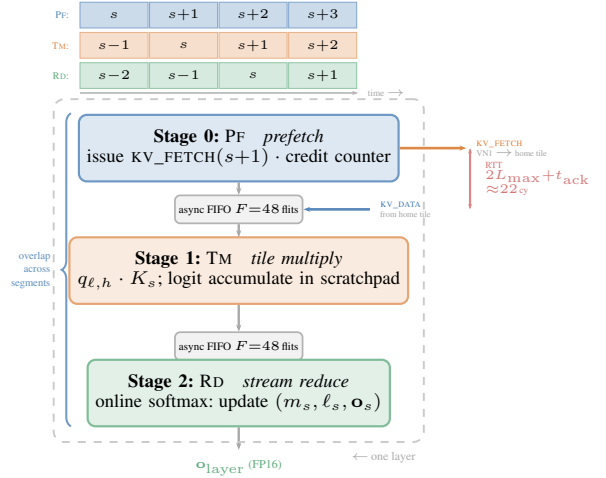

\pad{} overlaps three stages across segments
$s = 0, \ldots, \lceil T/P \rceil{-}1$ (Figure~\ref{fig:pad-pipeline}).
\textsc{Pf} issues a \textsc{KV\_FETCH} for $s{+}1$ iff the home injection
credit $C_{\pi}>0$ (one outstanding per $(\mathrm{seg},\ell)$).
\textsc{Tm} writes segment logits
\begin{equation}
  z_s = q_{\ell,h}\,K_{\ell,h,s}^{\top}/\sqrt{d_h}
  \label{eq:logit}
\end{equation}
into scratchpad SRAM ($\ge$80\% MAC when the FIFO is non-empty).
FIFO depth $F{=}48$ covers RTT $2L_{\max}{+}t_{\mathrm{ack}}{\approx}22$ cycles
plus $W_{\mathrm{coal}}$.
\textsc{Rd} applies FlashAttention-2 online softmax~\cite{dao2022flashattention,dao2023flashattention2}
\begin{align}
  m_s &= \max(m_{s-1},\,\mathrm{rowmax}(z_s)), \nonumber\\
  \ell_s &= e^{m_{s-1}-m_s}\ell_{s-1}
                 + \mathrm{rowsum}(e^{z_s-m_s}), \nonumber\\
  \mathbf{o}_s &= e^{m_{s-1}-m_s}\mathbf{o}_{s-1}
                 + e^{z_s-m_s} V_{\ell,h,s},
  \label{eq:online}
\end{align}
so $V$-flits never materialize a full attention matrix.

\section{Implementation}
\label{sec:impl}

We implement \mesh{} on a tiled FPGA accelerator: an
$8{\times}8$ interconnect with 512-bit wormhole links, four-bank
tile-local \kv{} SRAM, and $d_h{=}128$ systolic PEs, with \takv{},
\mare{}, and \pad{} in the hardware datapath.

We evaluate four configurations on the same platform:
\textsc{Cent}, which routes \kv{} traffic through a centralized memory path;
\textsc{Shared}, which distributes blocks with round-robin placement and
unicast; \takv{}-only, which enables affine placement without multicast; and
full \mesh{}.
Workloads are LLaMA-2-7B~\cite{touvron2023llama2} and
Mistral-7B~\cite{jiang2023mistral} at $T \in \{8192,32768\}$ and
$B \in \{1,4,8\}$.

Latency, throughput, traffic, and utilization are hardware measurements
(95\% CIs ${<}3\%$ throughput over five runs).
Per-router buffers plus a 256-bit bloom and a 16-entry exact tag table
(Table~\ref{tab:router}) require approximately one 18-Kb BRAM equivalent
per router.

\begin{table}[t]
\centering
\caption{Per-router parameters in the FPGA implementation.}
\label{tab:router}
\small
\begin{tabular}{@{}lc@{}}
\toprule
\textbf{Parameter} & \textbf{Value} \\
\midrule
Cardinal ports + tile injection & 5 \\
Flit width & 512\,bit \\
Unicast VCs (VN0)               & 2 \\
Multicast VCs (VN1)             & 2 \\
Per-VC input FIFO depth         & 8 flits \\
Segment bloom width             & 256\,bit \\
In-flight exact tags            & 16 entries \\
Bloom refresh interval $W_b$    & 96\,cycles \\
Coalescing window $W_{\mathrm{coal}}$ & 12\,cycles \\
\textsc{Pad} Pf/Tm FIFO depth   & 48\,flits \\
\bottomrule
\end{tabular}
\end{table}

Table~\ref{tab:fpga} reports implemented router RTL, organized as 8 tiles
per quadrant, together with \textsc{PE} and SRAM resources on an
$8{\times}8$ Alveo U280 instance. DSP48E2 is the bottleneck; router, bloom,
and tag logic use ${<}5\%$ of LUTs.

\begin{table}[t]
\centering
\caption{Alveo U280 resource utilization for an $8{\times}8$ \mesh{}
  instance (64 tiles; one attention layer). Percentages vs.\ DS963 device totals.}
\label{tab:fpga}
\small
\begin{tabular}{@{}lrr@{}}
\toprule
\textbf{Resource} & \textbf{Used} & \textbf{Available (\%)} \\
\midrule
LUT                        &   613\,K  & 47\% \\
FF                         &   820\,K  & 31\% \\
BRAM36K (router + control) &   768     & 38\% \\
URAM (KV staging SRAM)     &   512     & 53\% \\
DSP48E2 (MAC arrays)       & 5\,760    & 64\% \\
\bottomrule
\end{tabular}
\end{table}

\section{Evaluation}
\label{sec:eval}

\subsection{Interconnect, throughput, and latency}
\label{sec:eval:main}

Table~\ref{tab:traffic} reports \noc{} traffic at $T{=}32\mathrm{K}$, $B{=}1$,
normalized to \textsc{Cent}.
Affine \takv{} placement reduces traffic to
$0.66\times$.
\mare{} then suppresses redundant head copies, and the full design reaches
$0.42\times$ ($\mathbf{58\%}$ reduction).
Mistral-7B reaches $0.40\times$ because sliding-window attention
localizes some fetches.

Table~\ref{tab:util} and Figure~\ref{fig:motiv-traffic} decompose bisection
link cycles.
\textsc{Cent} expends nearly half of those cycles on back-pressure
($19\%$ useful \kv{} transfer at $T{=}32\mathrm{K}$).
\mesh{} converts these links into useful transfer: utilization reaches
$\mathbf{61\%}$, a $\mathbf{2.1\times}$ improvement over \textsc{Shared}.

\begin{figure}[H]
\centering
\begin{tikzpicture}
\begin{axis}[
  width=0.38\textwidth,
  height=3.2cm,
  ybar,
  bar width=6pt,
  ymin=0,
  ymax=2.4,
  ylabel={Normalized throughput},
  symbolic x coords={$B{=}1$,$B{=}4$,$B{=}8$},
  xtick=data,
  legend style={at={(0.5,1.03)},anchor=south,legend columns=-1,font=\scriptsize},
  enlarge x limits=0.25,
  grid=major, grid style={dashed},
  area legend,
  every axis plot/.append style={draw=none},
]
\addplot[fill=meshGray!50]
  coordinates {($B{=}1$,1.00) ($B{=}4$,1.00) ($B{=}8$,1.00)};
\addplot[fill=meshBlue!60]
  coordinates {($B{=}1$,1.19) ($B{=}4$,1.29) ($B{=}8$,1.46)};
\addplot[fill=meshGreen!60]
  coordinates {($B{=}1$,1.35) ($B{=}4$,1.61) ($B{=}8$,1.90)};
\legend{\textsc{Cent}, \textsc{Shared}, \mesh}
\end{axis}
\end{tikzpicture}
\caption{Decode throughput speedup at $T{=}32\mathrm{K}$, LLaMA-2-7B.}
\label{fig:tp_speedup}
\end{figure}
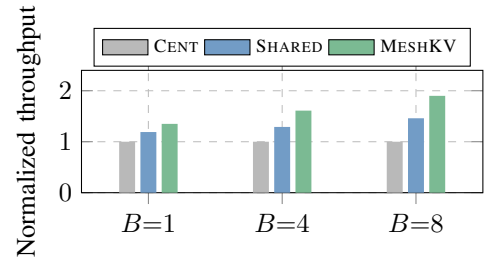

Figure~\ref{fig:tp_speedup} reports decode throughput at $T{=}32\mathrm{K}$.
At $B{=}1$ the speedup is $1.35\times$ from shorter hops and
fewer credit stalls.
At $B{=}8$, streams share segment multicast, and the speedup
increases to $\mathbf{1.90\times}$.

\begin{figure}[H]
\centering
\begin{tikzpicture}
\begin{axis}[
  width=0.38\textwidth,
  height=2.8cm,
  xlabel={Context $T$},
  ylabel={Latency ($\mu$s)},
  xmin=2000, xmax=34000,
  ymin=0, ymax=22,
  xtick={4096,8192,16384,32768},
  xticklabels={4K,8K,16K,32K},
  legend style={at={(0.5,1.03)},anchor=south,legend columns=3,
                font=\scriptsize,draw=none},
  grid=major, grid style={dashed},
  mark size=1.4pt,
]
\addplot[meshGray,thick,mark=square*]
  coordinates {(4096,4.2)(8192,9.3)(16384,16.9)(32768,21.6)};
\addlegendentry{\textsc{Cent}}
\addplot[meshBlue,thick,mark=triangle*]
  coordinates {(4096,3.9)(8192,7.3)(16384,11.6)(32768,16.5)};
\addlegendentry{\textsc{Shared}}
\addplot[meshGreen,thick,mark=*]
  coordinates {(4096,3.2)(8192,6.1)(16384,8.8)(32768,12.2)};
\addlegendentry{\mesh}
\end{axis}
\end{tikzpicture}
\caption{Latency per token vs.\ context length ($B{=}1$).}
\label{fig:lat_scaling}
\end{figure}
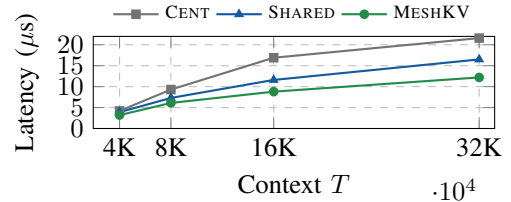

Fig.~\ref{fig:lat_scaling} reports per-token latency versus context length
(LLaMA-2-7B, $B{=}1$).
\textsc{Cent}'s injection queue saturates near 8K tokens, after which
latency grows steeply.
Under \mesh{}, latency remains approximately linear in $T$: \takv{}
distributes load and \mare{} reduces duplicate traversals.

\begin{table}[t]
\centering
\caption{Normalized interconnect traffic (\textsc{Cent}$\,{=}\,1.00$),
  $T{=}32\mathrm{K}$, $B{=}1$.}
\label{tab:traffic}
\small
\begin{tabular}{@{}lcccc@{}}
\toprule
\textbf{Model} &
\textbf{\textsc{Cent}} &
\textbf{\textsc{Shared}} &
\textbf{\takv{}-only} &
\textbf{\mesh{}} \\
\midrule
LLaMA-2-7B  & 1.00 & 0.81 & 0.66 & \textbf{0.42} \\
Mistral-7B  & 1.00 & 0.79 & 0.63 & \textbf{0.40} \\
\bottomrule
\end{tabular}
\end{table}

\begin{table}[t]
\centering
\caption{Bisection \kv{} bandwidth utilization (\%), LLaMA-2-7B.}
\label{tab:util}
\small
\begin{tabular}{@{}lcc@{}}
\toprule
\textbf{Configuration} &
\textbf{$T{=}8\mathrm{K}$} &
\textbf{$T{=}32\mathrm{K}$} \\
\midrule
\textsc{Cent}           & 29 & 19 \\
\textsc{Shared}         & 38 & 29 \\
\takv{}-only            & 52 & 41 \\
\mesh{} (full)          & \textbf{71} & \textbf{61} \\
\bottomrule
\end{tabular}
\end{table}

\subsection{Energy efficiency}
\label{sec:eval:energy}

At $T{=}32\mathrm{K}$, \mesh{} reduces link energy by $48\%$ relative to
\textsc{Cent}: \takv{} shortens paths and \mare{} eliminates duplicate \kv{}
copies, in line with the $58\%$ traffic cut in Table~\ref{tab:traffic}.
SRAM energy increases $7\%$ because \takv{} reads from distributed
banks rather than one contiguous array.
Net energy per token decreases by $\mathbf{17\%}$ under the
activity-based estimate in \S\ref{sec:impl}.
Shared multicast further amortizes the link term at $B{>}1$, where
concurrent streams reuse the same \textsc{KV\_DATA} tree.

\subsection{Comparison with prior \textsc{FPGA} decoders}
\label{sec:eval:prior}

FPGA \textsc{LLM} accelerators
DFX~\cite{hong2022dfx} and FlightLLM~\cite{zeng2024flightllm} optimize
end-to-end execution and HBM/BRAM management;
CXL-SpecKV~\cite{liu2026cxl} speculates \kv{} on a disaggregated FPGA.
\mesh{} implements the on-chip communication substrate for distributed
\kv{} movement, not a full FFN-and-norm decoder.
Different platforms, mappings, memory systems, and datapaths therefore make
absolute latencies incomparable.
We report hardware measurements against \textsc{Cent} and \textsc{Shared}
on the same $8{\times}8$ configuration: $58\%$ less interconnect traffic,
$\mathbf{2.1\times}$ \kv{} bandwidth utilization over \textsc{Shared}, and
up to $\mathbf{1.9\times}$ throughput at $B{=}8$
(Tables~\ref{tab:traffic}--\ref{tab:util}, Figure~\ref{fig:tp_speedup}).

\subsection{Ablation and sensitivity}
\label{sec:eval:ablation}

Table~\ref{tab:ablation} ablates each mechanism at $T{=}32\mathrm{K}$,
$B{=}8$ on LLaMA-2-7B (normalized to full \mesh{}).
Disabling multicast (\mare{} unicast) causes the largest regression:
throughput falls to $0.62\times$ and traffic rises to $1.55\times$.
Removing duplicate suppression and \pad{} overlap yields $0.82\times$ and
$0.78\times$ throughput, respectively; adaptive \takv{} swaps cost
$11$--$15\%$ on skewed prefixes.
At $d_h{=}128$, $P{=}64$ balances
$T \bmod P$ stragglers and flit overhead.
An $8{\times}8$ mesh attains $1.9\times$ over \textsc{Cent}, versus
$1.4\times$ on a $16{\times}4$ mesh whose longer multicast trees add hop
latency.
$W_{\mathrm{coal}}{=}12$ cycles merges $78\%$ of heads at the home tile.

\begin{table}[H]
\centering
\caption{Ablation at $T{=}32\mathrm{K}$, $B{=}8$, LLaMA-2-7B.
  Mean$\pm$95\% CI over five independent hardware runs;
  normalized to \mesh{}.}
\label{tab:ablation}
\small
\begin{tabular}{@{}lcc@{}}
\toprule
\textbf{Variant} & \textbf{TP} & \textbf{Traffic} \\
\midrule
\mesh{}        & $1.00{\pm}0.018$ & $1.00{\pm}0.027$ \\
$-$bloom       & $0.82{\pm}0.023$ & $1.38{\pm}0.041$ \\
$-$\takv{}    & $0.88{\pm}0.016$ & $1.22{\pm}0.029$ \\
$-$\pad{}      & $0.78{\pm}0.025$ & $1.00{\pm}0.021$ \\
$-$\mare{}     & $0.62{\pm}0.014$ & $1.55{\pm}0.044$ \\
\textsc{Cent}  & $0.53{\pm}0.012$ & $2.38{\pm}0.068$ \\
\bottomrule
\end{tabular}
\end{table}

\section{Conclusion}
\label{sec:conclusion}

\mesh{} recasts decode \kv{} movement as packetized mesh flows via
\takv{} striping, \mare{} multicast, and credit-aligned \pad{}.
Hardware measurements on FPGA datapaths show up to 58\%
less interconnect traffic, $2.1\times$ \kv{} bandwidth utilization, and
$1.9\times$ multi-stream throughput relative to same-platform baselines.
As context lengths grow, \mesh{} is a promising
communication-aware \kv{} fabric for long-context decode on tiled accelerators.


\FloatBarrier

\bibliographystyle{IEEEtran}
\bibliography{references}

\balance

\end{document}